\documentclass[preprint,12pt]{aastex}
\usepackage{emulateapj5}

\newcommand{\ee}[1]{\mbox{${} \times 10^{#1}$}}
\newcommand{\eten}[1]{\mbox{$10^{#1}$}}
\newcommand{\av}{\mbox{$A_V$}} 
\newcommand{\lbol}{\mbox{$L_{bol}$}} 
\newcommand{\lsun}{\mbox{L$_\odot$}}
\newcommand{\msun}{\mbox{M$_\odot$}}
\newcommand{\tbn}{\tablenotemark}

\slugcomment{\footnotesize {\LaTeX}ed on \today}
\shorttitle{IRAC observations of Cha~II}
\shortauthors{Porras et al.}

\begin{document}


\title {The Spitzer c2d survey of Large, Nearby, Interstellar Clouds. V.
Chamaeleon~II Observed with IRAC}

\author{
Alicia Porras\altaffilmark{1,}\altaffilmark{2},
Jes K. J{\o}rgensen\altaffilmark{1},
Lori E. Allen\altaffilmark{1},
Neal J. Evans II\altaffilmark{3},
Tyler L. Bourke\altaffilmark{1},
Juan M. Alcal{\'a}\altaffilmark{4}
Michael M. Dunham\altaffilmark{3}
Geoffrey A.  Blake\altaffilmark{5},
Nicholas Chapman\altaffilmark{6},
Lucas Cieza\altaffilmark{3},
Paul M.  Harvey\altaffilmark{3},
Tracy L. Huard\altaffilmark{1},
David W. Koerner\altaffilmark{7},
Lee G. Mundy\altaffilmark{6},
Philip C. Myers\altaffilmark{1},
Deborah L. Padgett\altaffilmark{8},
Anneila I.  Sargent\altaffilmark{9},
Karl R.  Stapelfeldt\altaffilmark{10},
Peter Teuben\altaffilmark{6}
Ewine F. van Dishoeck\altaffilmark{11},
Zahed Wahhaj\altaffilmark{7,}\altaffilmark{12}
\& Kaisa E. Young\altaffilmark{3,}\altaffilmark{13}
}

\altaffiltext{1}{Harvard-Smithsonian Center for Astrophysics, 60
  Garden Street, Cambridge, MA 02138; aporras@cfa.harvard.edu, 
jjorgensen@cfa.harvard.edu, leallen@cfa.harvard.edu, tbourke@cfa.harvard.edu,
thuard@cfa.harvard.edu, pmyers@cfa.harvard.edu}

\altaffiltext{2}{Instituto Nacional de Astrof\'{\i}sica, \'Optica y 
Electr\'onica (INAOE), Tonantzintla, Pue., M\'exico; aporras@inaoep.mx}

\altaffiltext{3}{Department of Astronomy, University of Texas at Austin,
  1 University Station C1400, Austin, TX~78712; nje@astro.as.utexas.edu,
mdunham@astro.as.utexas.edu, lcieza@astro.as.utexas.edu, 
pmh@astro.as.utexas.edu}

\altaffiltext{4}{INAF-Osservatorio Astronomico di Capodimonte, via
Moiariello 16, 80131 Napoli, Italy; jmae@sun1.na.astro.it}

\altaffiltext{5}{Division of Geological and Planetary Sciences,
  MS~150-21, California Institute of Technology, Pasadena, CA~91125;
gab@gps.caltech.edu}

\altaffiltext{6}{Department of Astronomy, University of Maryland, College
  Park, MD~20742; chapman@astro.umd.edu, lgm@astro.umd.edu, 
teuben@astro.umd.edu}

\altaffiltext{7}{Department of Physics and Astronomy, Northern Arizona
  University, NAU~Box~6010, Flagstaff, AZ~86011-6010; david.koerner@nau.edu,
Zahed.wahhaj@nau.ed}

\altaffiltext{8}{Spitzer Science Center, MC~220-6, California
  Institute of Technology, Pasadena, CA~91125; dlp@ipac.caltech.edu}

\altaffiltext{9}{Division of Physics, Mathematics, and Astronomy,
  MS~105-24, California Institute of Technology, Pasadena, CA~91125;
afs@astro.caltech.edu}

\altaffiltext{10}{Jet Propulsion Laboratory, MS~183-900, California
  Institute of Technology, Pasadena, CA~91109; krs@exoplanet.jpl.nasa.gov}

\altaffiltext{11}{Leiden Observatory, PO Box 9513, NL~2300 RA Leiden,
  The Netherlands; ewine@strw.LeidenUniv.nl}

\altaffiltext{12}{Institute for Astronomy, 2680 Woodlawn Drive,  
Honolulu, HI~96822; wahhaj@ifa.hawaii.edu}

\altaffiltext{13}{Department of Physical Sciences, Nicholls State University,
Thibodaux, Louisiana 70301; kaisa.young@gmail.com}


\begin{abstract}

We present IRAC (3.6, 4.5, 5.8, and 8.0 \micron) observations of the 
Chamaeleon II molecular cloud. The observed area covers about
1 square degree defined by $A_V >2$. Analysis of the data in the
2005 c2d catalogs reveals
a small number of sources (40) with properties similar to those of young stellar
or substellar objects (YSOs). The surface density of these YSO candidates 
is low, and 
contamination by background galaxies appears to be substantial,
especially for sources classified as Class I or flat SED. 
We discuss this problem in some detail and conclude that very 
few of the candidate YSOs in early evolutionary stages are actually 
in the Cha~II cloud. Using a refined set of criteria, we define a
smaller, but more reliable, set of 24 YSO candidates.

\end{abstract}

\section{Introduction}\label{intro}

One of the goals of the Spitzer Legacy Program ``From Molecular Cores to
Planet-forming Disks'' (c2d) is to observe a sample of
five molecular clouds (Chamaeleon II, Lupus, Perseus, Serpens and Ophiuchus)
selected to represent a wide range of physical conditions such as cloud mass,
gas density, extent, structure, and internal velocity dispersion 
\citep{evans03}.
The comparison between star-forming environments
in these nearby molecular clouds through observations in the
infrared will provide a better picture of the interplay of
ambient conditions and local star-formation processes.

The molecular cloud complex in Chamaeleon (hereafter Cha)
has three major clouds: Cha~I, Cha~II, and Cha~III \citep{schwartz77}.
These clouds are located somewhat below the Galactic plane
[$-16 <{\it b} < -13$, see Fig. 7 in \citet{lepine94}].
CO observations have been presented by \citet{vilas94}
and  \citet{mizuno01}, while $^{13}$CO and C$^{18}$O data are discussed
by \citet{mizuno99}. Comparisons
of the CO column density and extinction were made by \citet{hayakawa01},
who characterized the clouds as being very close to virial equilibrium,
though \citet{boulanger98} argued that Cha~II has two velocity 
components that may not be bound.
The estimated virial mass of Cha~II is 1.7$\times$10$^3$ \msun,
which is consistent with mass estimates from 100 \micron\ extinction maps
(Boulanger et al. 1998).

Cha II  is at a distance of 178 $\pm$ 18 pc
(Whittet et al. 1997). Its galactic coordinates of $l = 303$, $b = -14$
place it considerably off the Galactic plane; its high ecliptic latitude
of $\beta = -60$ implies that asteroids will be rare. The directions
of Galactic and equatorial coordinates are very similar for Cha Il.

Some low-mass cores are present in the Cha clouds \citep{vilas94}.
These cores have lower column density than
typical cores in Taurus and Ophiuchus, and the star formation efficiency is 
lower as well \citep{vilas94}.  Previous studies indicated that
star formation activity decreases  from Cha~I to Cha~III, with
Cha~II intermediate.  Cha~II was chosen for the c2d survey
to exemplify clouds with modest star formation activity, based on
previous data. The goal was to determine if the much more sensitive
data from Spitzer would change this picture.

A number of low mass T Tauri stars have been observed in
the Cha II cloud.  Nine of the Cha~II T Tauri stars (Class II, 
or stars with disks) selected from their H$_{\alpha}$ emission,
have been observed at 1.3 mm \citep{henning93}, with 3 detected;
their spectral energy distributions can be fitted by stars with 
disks with radii of about 25 AU.
This sample includes the only intermediate mass star in the cloud,
the Herbig Ae/Be star IRAS 12496-7650, also known as DK Cha \citep{hughes91}.
There is also a string of Herbig-Haro objects,
HH~52, 53, and 54, which move towards the NE direction
(see \S \ref{hh54}).
The YSO population in Cha~II has been discussed by a number of
authors, including \citet{alcala00}, \citet{vuong01}, \citet{persi03}, 
and \citet{young05}. In particular, \citet{persi03} used ISOCAM
and JHK images to study the population of young stars in an area
centered about 23\arcmin\ to the east of our coverage.
A number of the Class II sources were observed
with the IRS instrument on the Spitzer Space Telescope and show signs
of grain growth \citep{kessler06}.

We describe observations of Cha~II from 3.6 to 8 \micron, along with
complementary data from other bands in \S \ref{obs}
and a description of data
reduction and photometry in \S \ref{reduce}.
The results are presented in \S \ref{results}, and the identification and 
classification of candidate young stellar objects (YSOc) are in \S \ref{ysos}. 
We discuss problems of background contamination in \S \ref{classi}
and present a smaller, less contaminated sample based on refined
classification criteria.
We compare our results to previous work in \S \ref{objects}
and summarize our results in \S \ref{summary}.
This is a companion paper to one on c2d observations of Cha~II
at longer wavelengths \citep{young05}. 
It follows roughly the same plan as previous papers
on IRAC data from the c2d survey of Serpens \citep{harvey06serpens} and 
Perseus \citep{jes06},
in an attempt to provide a uniform set of statistics and diagnostic tools.

This paper does not present a complete list of YSOs associated with
Cha~II.
Analysis of the combined set of IRAC and MIPS data, along with data from
other studies, will be presented by Alcal{\'a} et al. (in prep.). That study
will include a complete list of YSOs, based both on our data and on work
that covers areas outside our survey (e.g., \citealt{persi03}).


\section{Observations}\label{obs}

\begin{figure}
\includegraphics[scale=0.80]{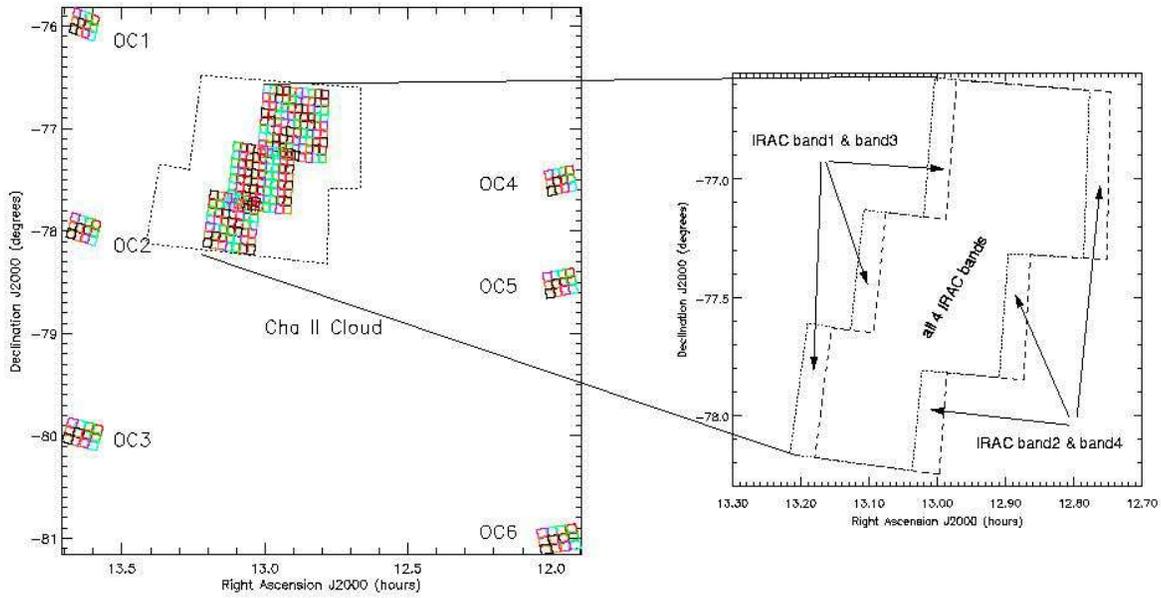}
\figcaption{\label{fig1_cov} {\it On the left.} Location of the IRAC coverage
maps of Cha~II, both on-cloud and the 6 fields off-cloud (OC). The smallest
squares represent
the IRAC FOV of $\sim$5\arcmin$\times$5\arcmin. Dotted line around the Cha~II
Cloud coverage sketches the observed area by MIPS (24$\mu$m) 
\citep{young05}.
{\it On the right.} Zoom that shows the areas surveyed by IRAC. Bands 1 and 3
boundaries are plotted with a dotted line, while bands 2 and 4 (slightly
shifted to the West), are plotted with a dashed line. 
Small areas observed in only two bands are shown with arrows. 
The observations lying in the greater central region covered by all the 4 bands
(band1=3.6$\mu$m, band2=4.5$\mu$m, band3=5.8$\mu$m, and band4=8.0$\mu$m), are
presented in this article.}
\end{figure}

\begin{figure}
\resizebox{\hsize}{!}{\includegraphics{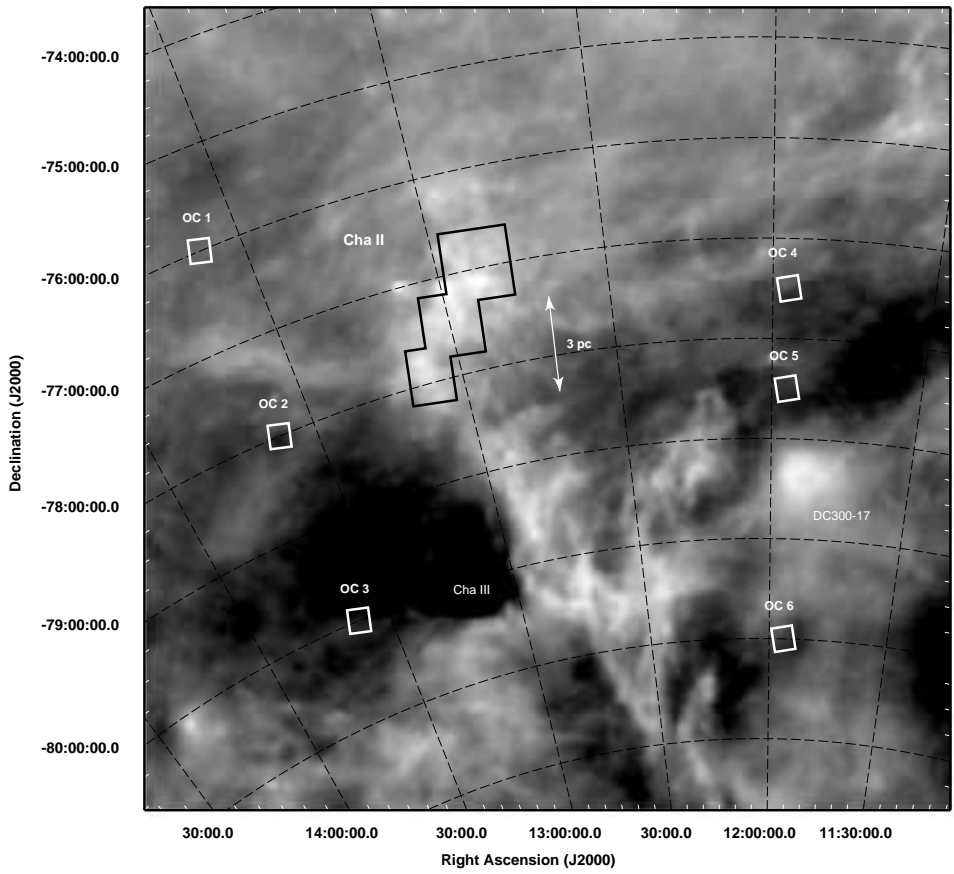}}
\caption{The coverage of the IRAC observations and the off-cloud fields
are shown superimposed on a gray scale IRAS map of emission at 100 \micron.
Some other nearby clouds are noted. The IRAC region indicated is the
region where all four bands overlap.
}
\label{irasfig}
\end{figure}

The Cha~II molecular cloud was observed with the Spitzer Space Telescope
(hereafter Spitzer) \citep{werner04}  with the Infrared Array
Camera (IRAC) described by \citet{fazio04}.  
We observed in the four IRAC bands
(with filters centered at about 3.6, 4.5, 5.8, and
8.0 $\mu$m) on 4 April 2004. The coverage is shown in detail in
Figure \ref{fig1_cov}.
The area mapped in these four bands
is shown on top of the IRAS 100 \micron\ image in Figure \ref{irasfig}.
It was selected based on the criterion that $A_V > 2$ mag
in maps made by \citet{cambresy99}, as shown in Figure 1 of \citet{evans03}.

The entire area was observed twice. The first observation, or epoch,
was separated from the second epoch by about 6.5 hours.
The first epoch was observed in high dynamic range mode (HDR, alternate short
and long time exposures) while the second epoch was observed 
in full array mode (FA, only one initial short time exposure
followed by several long time exposures).
The integration time per pointing, or frame time, was selected to be 12 sec.
HDR mode enables short integration time observations of 0.6 sec for
complementary
photometry of very bright sources that might saturate in the long exposure time
observations of 12 sec.
Every pointing in both epochs was repeated twice with a small positional shift
of $\sim$20-30$\arcsec$.
The combination of these two dithered images per pointing
not only increases the signal-to-noise of the observations but also facilitates
the removal of
some occasional bad pixels in the detector, as well as cosmic ray rejection.
The exposure time (i.e., the effective integration time per pixel) was
10.4 sec and 0.4 sec for the long and short integrations, respectively.
Thus, the total integration time of the combined epochs, with two dithers each,
is at least $\sim$42 sec and 0.4 sec (HDR) throughout the surveyed region.

The observed Cha~II cloud area was covered by three overlapping grids of
9 $\times$ 10, 8 $\times$ 9, and 7 $\times$ 7 pointings located in the northern,
middle, and southern parts of the cloud, respectively (Table 1).
Each individual IRAC pointing
has a FOV of $\sim5\arcmin \times 5\arcmin$ (see the coverage scheme in
Figure~\ref{fig1_cov}); thus the complete
mosaic of 211 pointings embraces more than one square degree, for each pair of 
bands, with the coverage for bands 1 and 3 offset to the east from that of
bands 2 and 4 (Fig. \ref{fig1_cov}).
The area covered by all four IRAC bands was $1.04$ 
square degrees or 10.0 pc$^2$. 

Six off-cloud (OC) regions were also observed with IRAC to compare the stellar
population found outside the molecular cloud region. These were selected
to have $\av < 0.5$ mag and little or no CO emission \citep{evans03}.
The coordinates of the
central position of these
OC regions and the three in the Cha~II cloud are summarized in 
Table 1, and their locations are shown in 
Figure \ref{fig1_cov}.
All six OC regions were covered by grids of 3 $\times$ 4 pointings. The
observations were made in HDR mode with
two repetitions on every two dither positions, and the frame time was 12 sec.
This parameter selection gives the same integration time for the off-cloud 
fields as for the on-cloud area.
The locations of the off-cloud regions are shown in Figure \ref{irasfig}.

All Cha~II IRAC observations are available at the SSC archive under the PID
number 176,
as part of the IRAC campaign 6. The AOR keys for Cha~II in-cloud are:
0005739520, 0005739776,
0005740032, 0005740288, 0005740544, and 0005740800, while for the OC regions:
0005741312,
0005741568, 0005741824, 0005742080, 0005742336, and 0005742592.

\section{Data Reduction}\label{reduce}

A detailed description of the c2d  pipeline can be found in the data delivery 
document\footnote{http://ssc.spitzer.caltech.edu/legacy/}
\citep{delivery3},
and a substantial description has been published by \citet{harvey06serpens}.
The source extraction and photometry will be described by Harvey et al.
(in prep.).
Consequently, we restrict ourselves here to
a summary that highlights points needed for later discussion. 

IRAC images were processed by the Spitzer Science Center (SSC)
using a standard pipeline (S11) to produce Basic Calibrated Data (BCD)
images. The data were then processed through the standard c2d pipeline.
The BCD images were
corrected by the c2d Calibration and Correction (CC) team for
some instrumental signatures providing Improved Calibrated Data
(ICD) to be mosaiced in every IRAC band. Sources were identified
using the mosaics, but the individual images were used for the final source 
extraction and photometry. The short exposure data were used for
bright sources to avoid saturation. An outlier rejection algorithm
was used to remove asteroids, which should be rare at the
ecliptic latitude of Cha~II.
Sources that did not fit a point source profile
were identified; most of the analysis in this paper is restricted to 
point sources.  The uncertainty in the flux calibration
is conservatively estimated to be 15\%.
Apparent magnitudes were calculated using the 
posted\footnote{http://ssc.spitzer.caltech.edu/irac/calib/overview.html}
IRAC zero magnitude flux densities.

A catalog of band-merged data with data quality flags was produced.
The IRAC data were merged with MIPS data at 24 \micron\ \citep{young05}
and with the Two Micron All Sky Survey (2MASS) catalog \citep{cutri03}.
For a source to be band-merged across IRAC bands or with 2MASS, a position
match within 2\arcsec\ is required; for MIPS 24 \micron, the match must
be within 4\arcsec.

\section{Results}\label{results}

The results in this paper are based on the c2d data products delivered
to the Spitzer Science Center
in December 2005 (hereafter 2005 products) and described by \citet{delivery3}.

\subsection{Cloud Images}\label{images}

\begin{figure}
\includegraphics[]{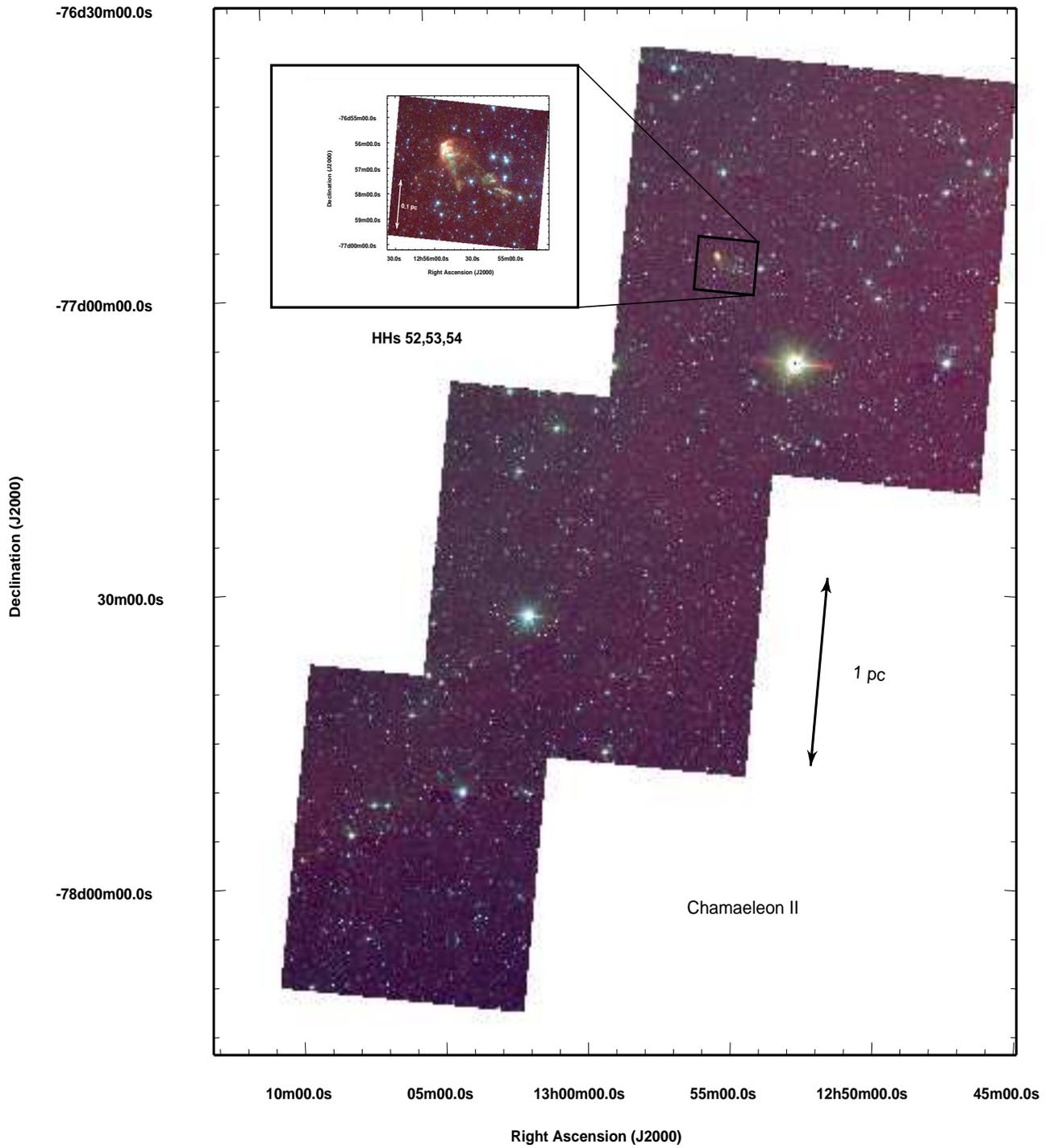}
\figcaption{\label{rgb1} Color composite image of the Cha~II molecular
cloud. Blue=3.6$\mu$m, green=4.5$\mu$m, and red=8.0$\mu$m. The total area
covered
by IRAC observations is about 1 square degree. Note the sparce distribution of
stars that characterizes this dark cloud.
The brightest mid-IR object corresponds to DK~Cha (IRAS 12496-7650), which is
related with the Herbig-Haro object HH~54 extended
towards the NE direction (see \S \ref{hh54})}
\end{figure}

\begin{figure}
\resizebox{\hsize}{!}{\includegraphics{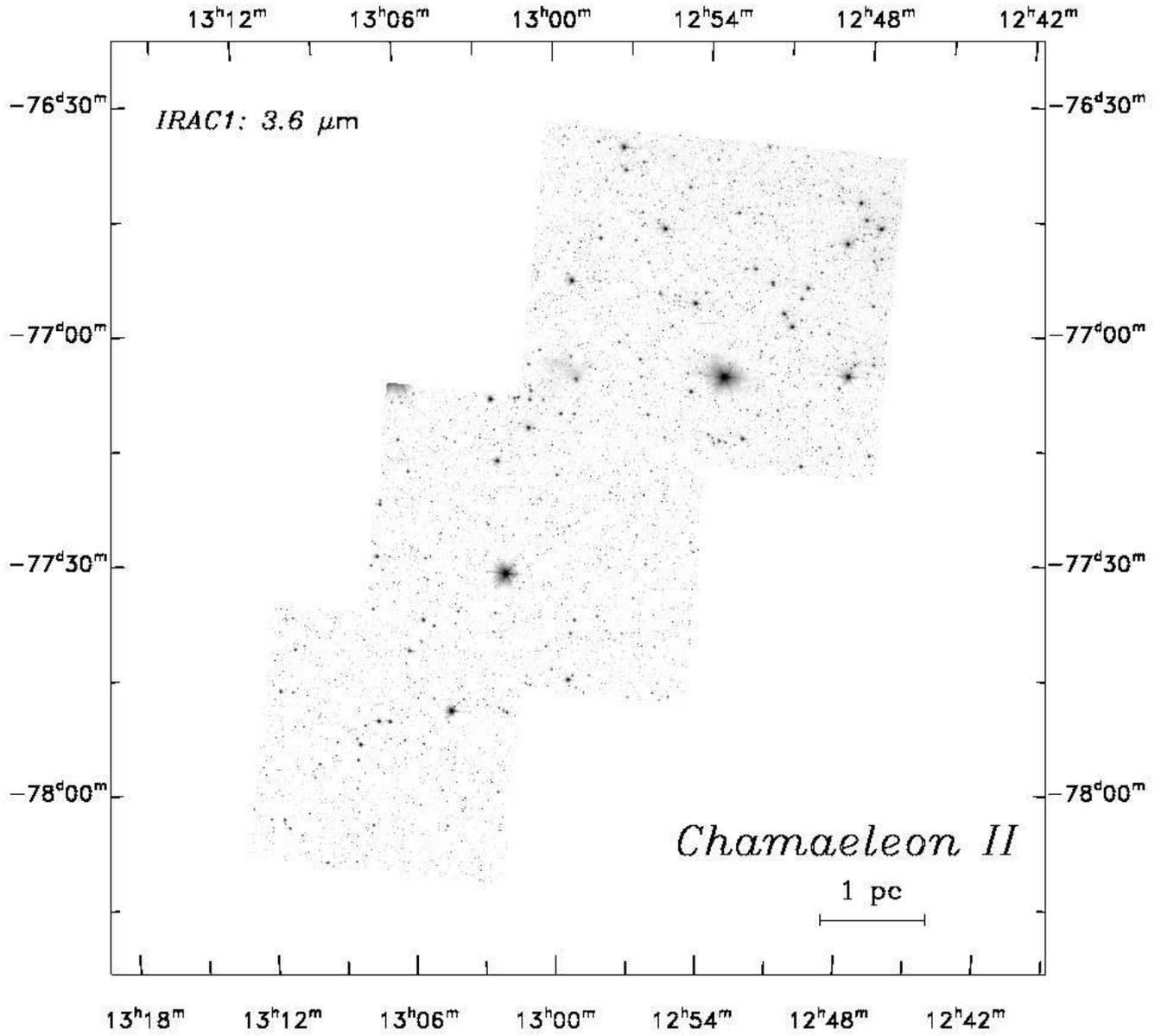}}
\caption{
Gray-scale images of the entire regions covered by each of the 4 IRAC bands. 
A logarithmic stretch was used for Bands 1 and 2, while a linear
stretch was used for Bands 3 and 4.
}
\label{fourbands}
\end{figure}

Figure \ref{rgb1} shows a color composite image in IRAC bands 1, 2, and 4
of the three overlapping regions of the  Cha~II molecular cloud. 
The gray-scale images for each band are shown in Figure \ref{fourbands}.
The brightest source corresponds to IRAS $12496-7650$, an Ae/Be star.
A Herbig-Haro object, HH~54, \citep{reipurth01}
is also visible in Figure \ref{rgb1} and
a blow-up of the region of the HH object is shown. This object will 
be discussed later (see section \ref{hh54}).

Compared to both Serpens \citep{harvey06serpens} and Perseus \citep{jes06},
Cha~II shows less diffuse emission at 8 \micron, indicative of a lower
level of interstellar radiation or a lower abundance of PAHs in Cha~II, or
both.
The MIPS images \citep{young05} show strong diffuse emission around 
IRAS~$12496-7650$ and in a region to the east of it, which unfortunately 
lies just off the edge of the IRAC images. 
This region, associated with the Class I source, IRAS $12553 - 7651$
(also known as ISO-ChaII-28)
is seen in the IRAS emission shown in Figure \ref{irasfig}.

\subsection{Source Statistics}\label{stats}

\begin{figure}
\resizebox{\hsize}{!}{\includegraphics{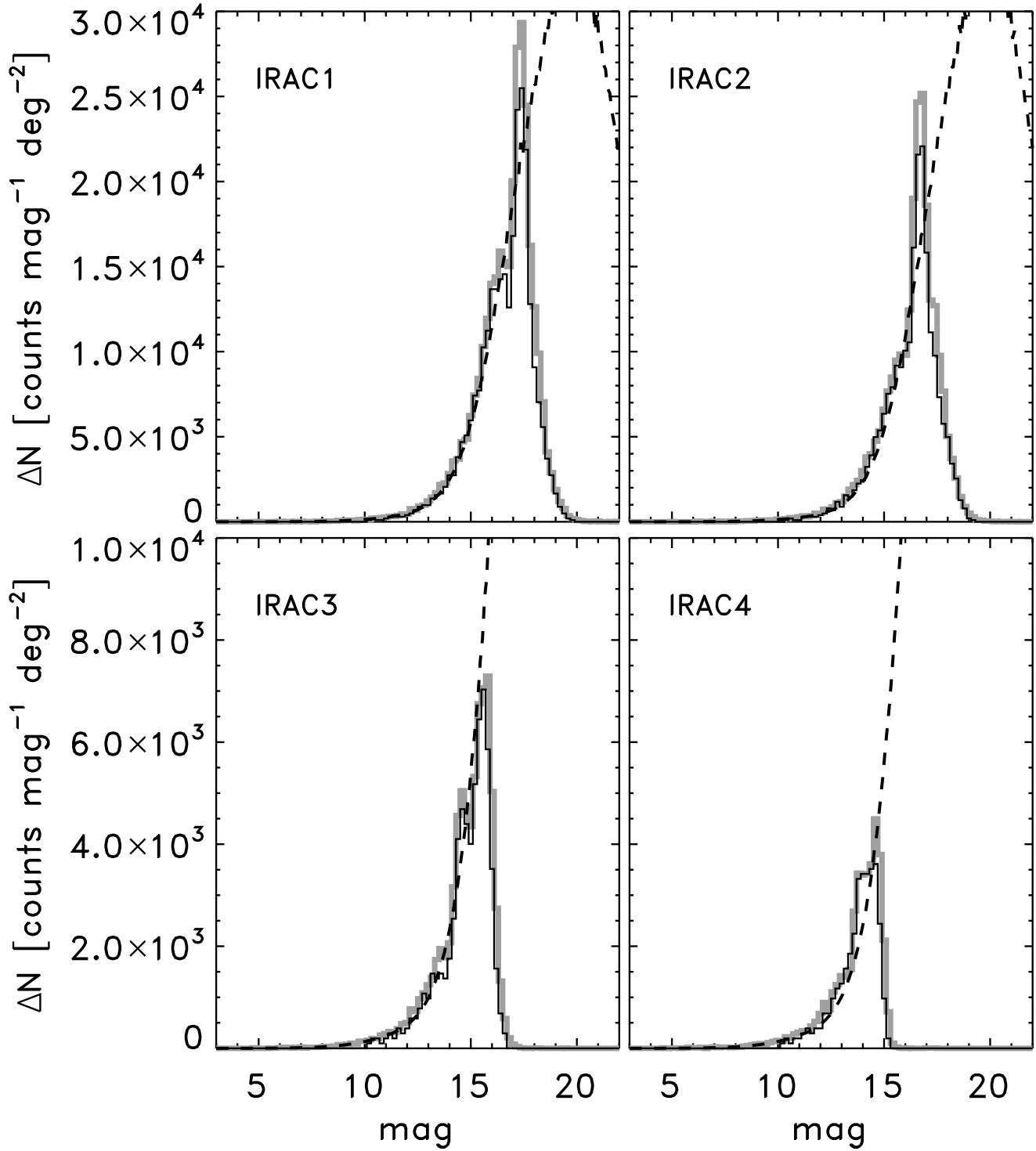}}
\caption{Differential source counts for the on- and off-cloud regions
  (grey and black, respectively). The predictions from the Wainscoat
  models are shown with the dashed line.} \label{diffcount}
\end{figure}

Table~\ref{detect1} shows the number of sources detected in Cha~II with
signal-to-noise of at least 7. The vast majority of the 69848 sources
detected in at least one band are background stars. This fact is illustrated
in Table~\ref{detect2}, where the number of sources per band is listed at
various levels of signal-to-noise. The number of sources detected at
each level is much greater in the shorter wavelength bands.
The differential source counts per solid angle for each band, shown in 
Figure~\ref{diffcount}, indicate very similar surface densities in the
on-cloud and off-cloud fields (gray and black lines, respectively).
The counts also agree closely with predictions of a Galactic model 
\citep{wainscoat92} until near the limiting magnitude, where an extra
peak is seen in both the on and off-cloud fields, most clearly in 
IRAC bands 1 and 3.
A similar effect was seen in Perseus \citep{jes06} and attributed to 
the distribution of background galaxies or stars.
The limiting magnitudes for a complete sample suggested by 
Figure \ref{diffcount} are 
17.5, 16.5, 15.0, and 14.0 mag for 3.6, 4.5, 5.8, and 8.0 $\mu$m respectively.
These are each about 0.5 mag higher than our pre-launch predictions for the
5-$\sigma$ sensitivities \citep{evans03}.

For further analysis, we restrict ourselves to the ``high quality"
catalog, which includes only sources with detections in all 4 IRAC bands,
excludes sources which were classified as extended during source extraction,
and excludes sources without a quality of detection of either ``A" or
``B" in at least one IRAC band (signal-to-noise greater than 
about 10 or 7 $\sigma$,
respectively). These criteria are discussed 
by \citet{harvey06serpens} and \citet{delivery3}; applying them leads to
a drastically pruned sample of 6284 sources (Table \ref{detect1}).
For comparison, 1532 sources were detected with MIPS at 24 \micron\ in the
somewhat larger field (1.5 square degrees) covered by that 
instrument \citep{young05}.  
The number of sources with 2MASS identifications and
detections in IRAC band 1 and band 2 is 4621 (Table \ref{detect1}). 
Many on-cloud sources (2179) were detected by IRAC, 
but not 2MASS;
a few sources (67) were detected (S/N at least 10 in both H and K) by 2MASS,
but not IRAC. Such sources are generally located around the edges of the
IRAC coverage, though a few correspond to sources that saturated the IRAC
images. The two brightest sources are not listed in our high-quality catalog
because they are saturated; these are IRAS $12496-7650$ (DK Cha) and 
IRAS $12580-7716$ (CD$-76\ 565$).

In comparison to Perseus \citep{jes06}, Cha II has a similar number of
4-band detections in an area that is smaller by nearly a factor of 4;
this difference arises because Cha~II is somewhat closer to the Galactic
plane ($b = -14$) than Perseus ($b \approx -20$), and closer to the Galactic
center ($l = 303$ versus $l = 160$). The number of 4-band
detections in Serpens  ($b \approx 5$) is still greater than in Cha~II 
by a factor of more than 2 \citep{harvey06serpens}.

\section{Candidate Young Stellar Objects: the 2005 Catalog}\label{ysos}

The sensitivity of the c2d survey is sufficient to detect embedded sources
and young stars down to very low levels of luminosity ($L \sim 10^{-3}$ \lsun). 
We refer to such objects as Young Stellar Objects (YSOs) whether they are in 
fact stellar or substellar, and we include all stages of evolution from deeply
embedded protostars to revealed stars as long as they have a convincing
infrared excess over the stellar photosphere.

As was clear from the discussion in \S \ref{stats} and from Figure 
\ref{diffcount}, the vast majority of sources that we detect are background
stars, while background galaxies represent a significant contaminant at
low flux levels. The question is how to separate the wheat from the chaff.
As discussed by \citet{harvey06serpens} and \citet{jes06}, our
2005 catalogs \citep{delivery3} did not effect a complete separation. Instead,
we used a combination of criteria based on color-magnitude and color-color
diagrams to produce samples ``enriched" in YSOs and ``impoverished" in
background stars and galaxies. We will explore this sample, discuss its
limitations, and then describe a more refined sample  based on
an improved separation procedure that is currently being tested.

\subsection{Diagnostic Diagrams}\label{DDs}

\begin{figure}
\resizebox{\hsize}{!}{\includegraphics{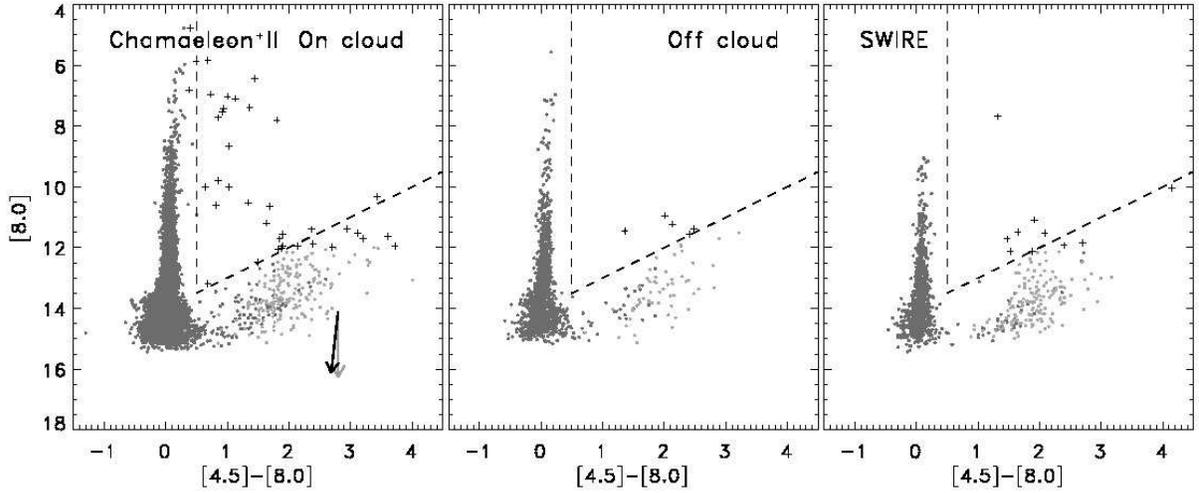}}
\caption{$[8.0]$ vs. $[4.5]-[8.0]$ color-magnitude diagrams based on
  the high quality catalogs (see text) with the Cha~II ``on-cloud''
  regions (left), ``off-cloud'' region (middle) and resampled SWIRE
  catalog (right). In each panel stars have been indicated by dark
  grey dots, YSO candidates by black plus signs and other sources by
  light grey dots. The distinction between YSO candidates and other
  sources was predominantly made based on this diagram with YSO
  candidates lying between 
  the two dashed lines with a few additional sources added
  based on their $[8.0]-[24]$ colors (see discussion in text). Two
  extinction vectors are shown corresponding to $A_K$ = 5 mag. The
  gray vector was derived for diffuse ISM regions by Indebetouw et
  al. (2005).  The black vector, appropriate for the dense regions
  found in molecular clouds and cores, was derived from deep
  near-infrared and Spitzer observations of dense cores Huard et
  al. (in prep.).}\label{i24_v4}
\end{figure}

\begin{figure}
\resizebox{\hsize}{!}{\includegraphics{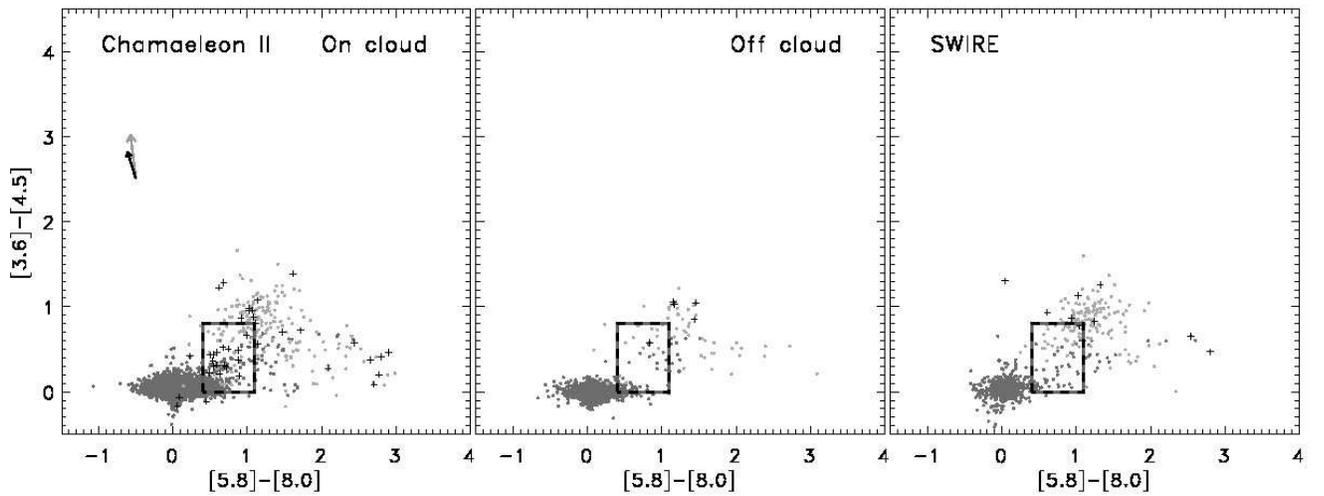}}
\caption{$[3.6]-[4.5]$ vs. $[5.8]-[8.0]$ color-color diagram for the
  ``on-cloud'' (left), ``off-cloud'' (middle) and SWIRE (right)
  fields. Sources, symbols and extinction vectors as in
  Fig.~\ref{i24_v4}. Class II sources generally lie within the box
indicated on the figures, while Class I sources lie above or to the
right of the box \citep{allen04}. Galaxies clearly are more likely to
contaminate the Class I region of color-color space.
}\label{i34_v12}
\end{figure}

\begin{figure}
\resizebox{\hsize}{!}{\includegraphics{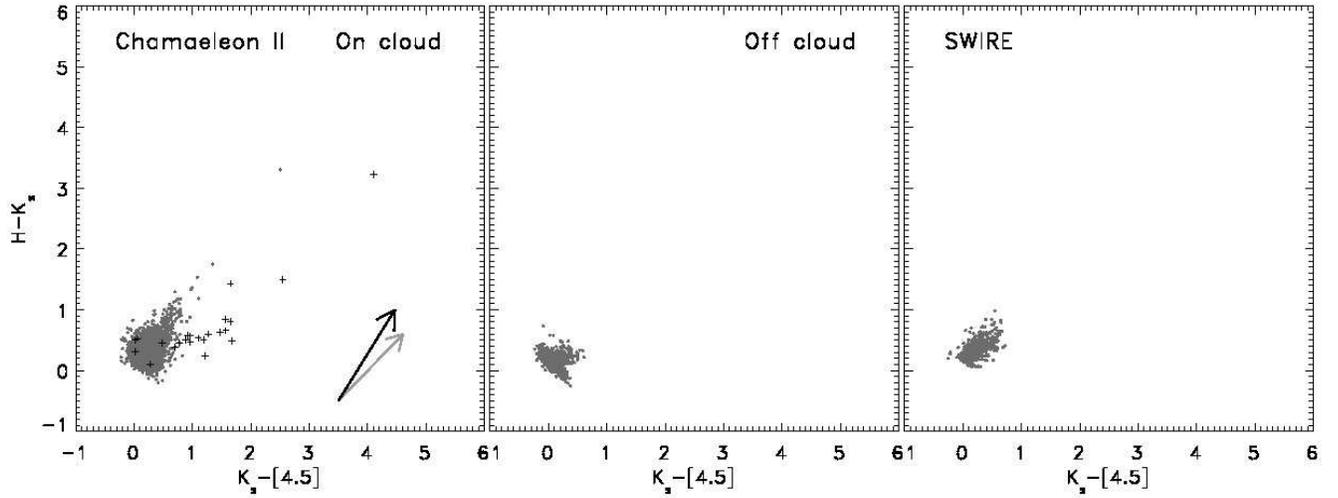}}
\caption{As in Fig.~\ref{i24_v4} and \ref{i34_v12} but for a
  $H$-$K_s$ vs. $K_s$-[4.5] color-color diagram. All sources with high
  quality detections in the IRAC bands 1 and 2 (see text) and
  10$\sigma$ detections in $H$ and $K_s$ included. Otherwise sources
  and symbols as in Fig.~\ref{i24_v4} and \ref{i34_v12}; extinction
  vectors shown are for $A_K=2$~mag.}\label{ihk_vk2}
\end{figure}

Figures \ref{i24_v4} to \ref{ihk_vk2} show color-magnitude or color-color
diagrams for Cha~II, the off-cloud control region, and data from the 
SWIRE observations \citep{surace04} of the ELAIS N1 field \citep{rowan04}. 
Only the ``high-quality" sources,
as defined above, are included. The SWIRE field should not include
any YSOs and should represent the background galaxy distribution. The
SWIRE data were processed through our pipeline in the same way as the c2d data
for consistency. Then our SWIRE catalog was statistically extincted with
the same distribution of extinction levels seen toward Cha II and 
resampled to best match the limiting magnitudes of our Cha~II observations 
(P. Harvey, in prep.). Several independent samples were made of 
the full 5.3 deg$^2$ area to check for variations, which were small.
For the figures, the sources from one of these samples 
were again randomly sampled to simulate the solid angle covered by our
Cha II data.

Figure \ref{i24_v4} shows a color-magnitude diagram for 
Cha~II, the off-cloud control region, and the resampled version of 
the SWIRE sample. Stars with no infrared excess occupy a band around
zero color in $[4.5]-[8.0]$; sources to the right of the vertical dashed
line at $[4.5]-[8.0] > 0.5$ have significant infrared excess. This criterion
is based on finding a minimum in the distributions of stars and galaxies from
the SWIRE sample at this color (S. Lai et al., in prep.).
At faint 8.0 \micron\ magnitudes,
a second distinct clump of faint sources appears. This clump is seen in
both the off-cloud field and the SWIRE data as well, and it is clearly
composed primarily of background galaxies. Analysis of the SWIRE data
indicates that 95\% of the galaxies lie below a line corresponding to
$[8.0] = 14-([4.5]-[8.0])$, as shown by the diagonal dashed line. 
In the 2005 catalog, we classified
sources lying between the two dashed lines as ``YSO candidates" (YSOc),
as discussed in \citet{delivery3} and S. Lai et al. (in prep.),
and these are indicated by plus signs in Figure \ref{i24_v4}.

Some sources outside the dashed lines are also classified as YSOc; these
have satisfied a similar set of criteria in another color-magnitude plot
using the MIPS 24 \micron\ data. That criterion typically finds stars
with infrared excesses that are significant only for $\lambda > 8$ \micron;
typically it provides only a modest increment to the YSOc category ($\sim 10$
sources out of the total of 40 YSOc in the case of Cha~II).
Some relatively bright galaxies will be misclassified as YSOc in this process;
based on the resampled SWIRE sample, there may be $\sim 17$ such galaxies in the
Cha~II sample, but extinction in the cloud might make this an upper limit.
The extinction vectors in Figure \ref{i24_v4} show that extinction
will tend to move objects below the diagonal line. It is also possible that we
are excluding real YSOs from the YSOc category because they are faint or
highly extincted. The Serpens data show evidence for a population of
such faint sources \citep{harvey06serpens}. 

Figure~\ref{i34_v12} shows the color-color diagram with $[3.6]-[4.5]$ vs. 
$[5.8]-[8.0]$ (\citealt{megeath04}; \citealt{allen04}), 
and Figure~\ref{ihk_vk2} shows the
$H-K_s$ vs. $K_s-[4.5]$ \citep{gutermuth04} color-color diagram. These
diagrams have proven useful in separating YSOs from stars, and our
classification of YSOc (again plotted as plus signs) is generally consistent.
Clearly, the galaxies cannot be excluded from the  $[3.6]-[4.5]$ vs. 
$[5.8]-[8.0]$ diagram, as many occupy the same space as the YSOs. The
$H-K_s$ vs. $K_s-[4.5]$ diagram is much more effective in separating galaxies,
but the requirement for data at $H$ and $K_s$ might eliminate a substantial
number of YSOc sources, especially the most deeply embedded, as long as
only 2MASS data are available. For Cha~II, only 22 of the 40 YSOc
can be plotted. This diagram illustrates how deeper 
near-infrared surveys could effect a cleaner separation of the YSOs 
and galaxies. However, extinction can have a bigger impact on this
diagram, as shown by the arrows.

\subsection{Classification of YSO Candidates}\label{yso_cc}

Using the sample of YSO candidates selected in \S \ref{DDs},
we characterize sources in Cha~II. In doing so,
we must bear in mind that this sample only purports to be
{\it enriched} in actual YSOs. We will explore the effects of contamination
in the next section.

Table~\ref{ysotab} shows the number of YSO candidates and divides them into
classes. Both the total number (40) and the number per square degree (38.5)
are much less than what was found toward Serpens 
($N = 257$, $dN/d\Omega = 289$) by \citet{harvey06serpens} 
and Perseus ($N = 400$, $dN/d\Omega = 104$) by \citet{jes06}.
Moreover, this calculation does not account for possible
contamination by background galaxies.

The sources were divided into the traditional classes \citep{lada87}
based on a fit to their spectral slope: 
$\alpha=\frac{{\rm d}\log \lambda F_\lambda}{{\rm d}\log\lambda}$.
We use a least-squared fit to all photometric points available 
between $K_s$ and 24 \micron\ in the calculation of $\alpha$. 
We then place these in the class system as extended by \citet{greene94}:
\begin{description}
\item[I] $0.3 \leq \alpha $
\item[flat] $-0.3 \leq \alpha < 0.3$
\item[II] $-1.6 \leq \alpha < -0.3$
\item[III] $\alpha < -1.6$.
\end{description}

We have arbitrarily included sources with $\alpha = 0.3$ in Class I. 
As noted in column 2 of Table~\ref{ysotab}, Cha~II would appear to host 
roughly equal numbers of
Class I and flat sources versus Class II sources, with a smaller number
of Class III sources. Since the Class III category does not require any
minimum value of $\alpha$, while we include only sources with significant
infrared excesses, the low counts in Class III should not be over-interpreted.
Data at other wavelengths are needed to add pre-main-sequence stars without
substantial infrared excess.

The ratio of Class I and flat spectrum to Class II sources in Cha~II
(1.1) is higher than in either Serpens (0.39) or Perseus (0.51), suggesting
that star formation in Cha~II is at a relatively early stage. The ratio
is comparable to that found in the extended Perseus cloud, after the
clusters (IC348 and NGC1333) are removed. 
This would be a surprising result,
and we examine it critically in \S \ref{classi}.

\begin{figure}
\includegraphics[scale=0.6]{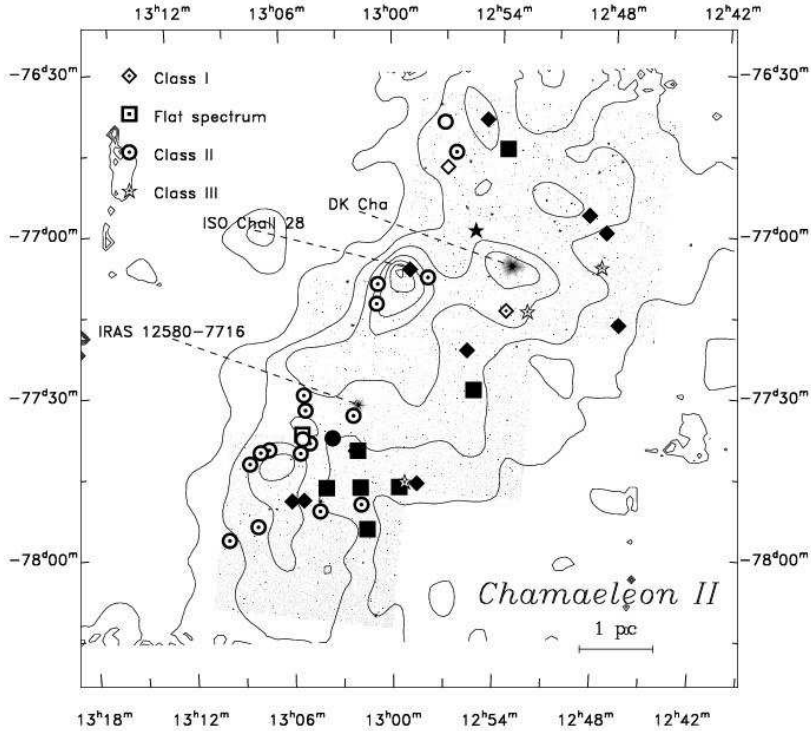}
\figcaption{\label{spa_dist} Spatial distributions of sources in
different classes in the Cha~II molecular cloud. These are represented
by different symbols plotted over an image of the 4.5 \micron\
emission and contours of visual extinction, with $\av = 2$ up to 12
in steps of 2 mag. The extinctions are based on fits to extinction toward
sources identified as stars. Two objects that are too bright to appear
in our catalog are indicated by name. The solid symbols indicate objects
that were classified as YSOc in the 2005 Catalog, but that are rejected
by the new criteria. Most of the rejected sources lie in regions of low 
extinction. The exception is ISO-ChaII-28, which lies near the peak of
extinction east of DK Cha; as noted in the text, this is a legitimate YSO.
The three new YSOc selected by the new criteria are shown without central
dots. The interacting galaxies are the open ``Class I" symbol near the top.
The two legitimate YSOs are open ``Class II" symbols near the top and in
the cluster around the southern extinction peak.
}
\end{figure}

The spatial distribution of YSOc sources by class in Cha~II 
(Fig.~\ref{spa_dist}) 
shows that in general, all classes are sparsely distributed in 
the molecular cloud, particularly in the middle of the central region. 
This is a region of relatively modest extinction.
There are small groups of Class II sources near extinction peaks. 

\section{Low Luminosity Class I Sources or Extragalactic Vermin?}\label{classi}

In this section, we reconsider the criteria for classifying sources
as YSO candidates. We will conclude that many of those classified as YSOc
in the 2005 catalog, especially those classified as Class I or flat SEDs,
are in fact extragalactic background sources. This discussion provides
a warning to people using those catalogs. We then discuss steps currently
underway to obtain a more reliable sample.

The statistics of the last section suggested a surprisingly large fraction of
Class I sources in the Cha~II cloud. Such an early phase of star formation
in this cloud would be at odds with previous studies and with the lack of
strong continuum emission at millimeter wavelengths over most of the cloud 
\citep{young05}.  Furthermore, most of the putative
Class I sources have very low luminosity ($\lbol \sim \eten{-3}$ \lsun), after
integrating over all wavelengths with {\it available} photometry from 1.2 to 
70 \micron. (Most are not detected at $\lambda < 3.6$ \micron.)
 Such low apparent luminosities for Class I sources, if real,  would
be very interesting in light of the emerging evidence for a class of
Very Low Luminosity Objects (VeLLOs) in some of the small clouds in
our study (\citealt{young04}; \citealt{bourke06}; \citealt{dunham06}),
though the established VeLLOs have $\lbol \sim 0.05$ to 0.1 \lsun,
rather than $\lbol \sim \eten{-3}$ \lsun. 

A comparison to the table of YSO candidates in \citet{young05}
found matches for only 2 of the 11 putative Class I sources. However, all but
two of our sources do have detections at MIPS 24 \micron; their absence from
the candidate list of \citet{young05} is caused by a lack of
2MASS detections, which was required by \citet{young05}.

The Cha~II cloud presents a good test case for checking for extragalactic
contamination: the surface density of sources is low, making such
contamination relatively more important; and the small number of sources
allows careful examination. We focus here on the sources classified as
Class I or flat SED, as these would be the most surprising. 
We will employ
methods that do not require careful examination of each object, so
that these methods may be useful for other clouds. 

We will first consider a statistical correction based on our 
extincted and resampled SWIRE data from the ELAIS-N1 field.
(The off-cloud data also suggest substantial contamination, but the statistics
are too poor to do a detailed correction.)
We use a section of the SWIRE data covering 5.3 deg$^2$ to obtain source
counts per solid angle. Then we normalize to the area covered in Cha~II.
We assume that all sources classified
as YSOc in the SWIRE sample are in fact background contamination. 
The number of sources classified as YSOc in the off-cloud and SWIRE
fields (also shown in Table \ref{ysotab}) indicates that about 12 of
the 40 YSOc sources in Cha~II could be background galaxies. The preponderance
of Class I and flat SED sources among these galaxies is clearly seen in those
entries in Table \ref{ysotab} and by the distribution of ``YSOc" in the
SWIRE panel of Fig. \ref{i34_v12}.

Focusing on the 19 putative Class I and flat SED sources, the statistics of the
SWIRE data suggest that only about 9 should actually be embedded sources.
Small number statistics make this number very uncertain. Can we decide
which are bona fide Class I sources and which are extragalactic vermin?
Only two have $L > 0.1$ \lsun; both of these are IRAS sources and they
are the two Class I sources also identified as YSOs by \citet{young05}.
Of the other 9 Class I sources, 7 have $L < 3\ee{-3}$ \lsun, assuming
the distance to Cha~II. These are very unlikely to be actual Class I sources.

We conclude that extragalactic vermin are a substantial source of fake 
Class I and flat SED sources in Cha~II. 
We caution that extragalactic vermin can be a significant source of false YSOc 
in early stages in fields with low surface densities of bona fide sources,
leading to distortions in source counts and classification schemes.

\subsection{Toward a Cleaner Sample of YSOs}\label{newysos}

We are currently working on a set of criteria that will effect a much
cleaner separation between true YSOs and background galaxies. These
criteria will be described in detail by Harvey et al. (in prep.).
They use both IRAC and MIPS color and magnitude criteria. 
When applied to the Cha~II data and the extincted, resampled SWIRE data,
these new criteria deliver much smaller numbers of YSO candidates:
24 in Cha~II and 3 in the full SWIRE sample covering 5.3 deg$^2$.
This would predict that only 0 or 1 fake YSO should appear in the
new sample of YSOc in Cha~II, which covers only 1.04 deg$^2$.

It is instructive to compare the old and new YSOc samples.
The 19 sources rejected by the new criteria are almost all the
sources of apparent low luminosity (14 of 19 with $L < 3\ee{-3}$ 
\lsun) and early (Class I or flat) SEDs (17 of 19). Only one reject
is a legitimate YSO, ISO-ChaII-28, or IRAS $12553-7651$ 
(\citealt{persi03}; \citealt{young05}).

The YSOc selected by the new criteria include 3 sources not included
previously. Two of these are legitimate YSOs based on other work:
Sz54 \citep{hughes92} and IRAS $12535-7623$ (J. Alcal{\'a} et al. in prep.). 
They were previously excluded by the apparent presence of a second source
within 2\farcs0 in the 4.5 \micron\ band. 
This criterion has been relaxed in the new criteria.
The third new source is an interesting case study in the remaining
problems. It was previously excluded because it was fitted as an extended
source in two bands. It is brighter at long wavelengths than any source in
the larger SWIRE field, and it has a steeply rising spectral index (1.66).
It would have a luminosity out to 70 \micron\ of about 0.05 \lsun.
It is either a VeLLO or an unusally bright background galaxy.
Fortuitously, this object has been observed with HST in a study of
putative edge-on disks (Stapelfeldt et al. in prep.). In that image,
it is clearly a pair of interacting galaxies!

The YSOc rejected by the new criteria are shown as filled symbols in 
Fig. \ref{spa_dist};
they mostly lie in regions of low extinction, though they are surprisingly
clustered. The surviving YSOc correlate better with the extinction contours.

The new criteria seem to be much more effective. They have included only
one clear background galaxy (pair) and they have excluded one
legitimate YSO, a much better record. If we delete the known contaminant
and add the known left-out source, we get the statistics in the
``New" Column in Table  \ref{ysotab}.
Recall that the two brightest sources are not included because of
saturation; of these, DK Cha would be a Class I YSO \citep{henning93}.
Further analysis of the entire YSO population 
requires data at other wavelengths and will be pursued
in a follow-up paper (J. Alcal\'a, in prep.). Early results from that work
support the value of the new criteria in separating YSOs from background
galaxies.

\section{Comparisons to Other Studies and Comments on 
Individual Objects}\label{objects}

The two certain Class I sources (IRAS $12500-7658$ and IRAS $12553-7651$,
also known as ISO-Cha II-28) have been discussed by \citet{young05}.
Rather than discuss individual sources in detail, we engage in some
comparison of our samples of YSOc to those found in previous studies
with the goal of illustrating the complementary nature of different
surveys.

The YSO population in Cha~II has been analyzed by
\citet{vuong01}, \citet{persi03}, \citet{alcala00},
and \citet{young05}. A number of brown dwarfs with disks have been
identified using deep near-infrared data and our c2d catalogs
\citep{allers06}. Some of the sources discussed
by these authors show up as YSOc sources in our catalogs, primarily
in the classes from ``flat" to Class II. We describe briefly comparisons
to some of the previous catalogs.

Six of our YSOc list are found in Table 1 of \citet{vuong01}, selected
on the basis that $I-J \geq 2$.
Only three of the objects in the Tables in \citet{persi03} are in our
YSOc list; the overlap is small because the area covered by their study
was smaller and about half is off to the east of our coverage.
Of the seven candidate brown dwarfs with disks in Cha~II listed
by \citet{allers06}, 5 appear in our YSOc list. The two missing sources
are too faint to be included in the 2MASS catalog, and hence have 
insufficient wavelength coverage to be classified in our catalogs, even
though they are well detected. \citet{allers06} were able to classify
them using deeper near-infrared data. Most of the objects found by
\citet{allers06} have now been confirmed with spectroscopy to
be brown dwarfs (\citealt{alcala06}; \citealt{allers06b};
\citealt{rayjay06}).

\subsection{HH~54}\label{hh54}

The three color image in Fig. 10 shows the Herbig-Haro objects HH 54 (upper
left) and HH 52/53 (lower right).  HH 54 is a well studied region notable
for its large abundance of water vapor \citep{liseau96} and
ratios of ortho-to-para H$_2$ below those expected for the observed gas
temperatures ($>2000$ K), as discussed by \citet{gredel94} and 
\citet{neufeld98}.  Observations
of HH~54 with IRS aboard Spitzer \citep{neufeld06} show that 
IRAC bands 3 and 4 are dominated by shocked rotational transitions of
H$_2$, tracing temperatures of 400-1200 K, while H$_2$ emission in IRAC
band 2 traces somewhat higher temperatures and is a good tracer of shocked
outflow emission without the possible confusion from PAH emission 
(\citealt{smith05}; \citealt{noriega-crespo04}).
The bow shock of HH 54 shows a
mix of green and red emission (IRAC bands 2 and 4 respectively), while the
region behind the bow and around HH 52/53 is dominated by green emission
(IRAC band 2 only), indicative of higher temperatures and in broad
agreement with the trends seen by \citet{neufeld06}.  HH 54 is also
well detected at 24 \micron\ by MIPS and IRS.

\section{Conclusions}\label{summary}

The Cha~II molecular cloud has been surveyed with the IRAC instrument
of the Spitzer Space Telescope. Among the nearly 70,000 sources detected
in at least one band in our 2005 catalogs, 40 objects have been identified 
as candidates to be
young stellar or substellar objects, collectively known as YSOc's.
The total number and number per solid angle of YSOc's are much smaller
than toward other c2d clouds analyzed so far, Serpens and Perseus.
The number of the YSOc's that were classifed as Class I or flat spectrum
sources is comparable to the number in the Class II stage, unlike the
situation in Serpens and Perseus. If real, these would indicate much
more active star formation than previously known in Cha~II, concentrated on
very low luminosity objects. However, many of these appear to be background
galaxies; improved criteria yield only 24 objects classified as YSOc, mostly
Class II objects.
Care must be taken to avoid distortion of the statistics of
sources in various classes by background objects, especially in fields
with low surface density of YSOs, such as Cha~II. The contaminating
background galaxies are preferentially found in the earlier classes.

\acknowledgments We thank the referee for suggestions which led to 
a clearer presentation. We are grateful to the staff at the Lorentz Center at
Leiden University for hospitality during a three week meeting in July
2005 where a large part of this work was pursued. 
Support for this work, part of the Spitzer Legacy Science
Program, was also provided by NASA through contracts 1224608, 1230782,
and 1230779 issued by the Jet Propulsion Laboratory, California
Institute of Technology, under NASA contract 1407. 
The research of JKJ was supported by NASA Origins Grant NAG5-13050. 
Astrochemistry in
Leiden is supported by a NWO Spinoza grant and a NOVA grant. KEY was
supported by NASA under Grant NGT5-50401 issued through the Office of
Space Science. This publication makes use of data products from the 
Two Micron All Sky Survey, which is a joint project of the University
of Massachusetts and the Infrared Processing and Analysis Center/California
Insitute of Technology, funded by NASA and NSF.
This research has made use of the SIMBAD database, operated at CDS, Strasbourg,
France.


\begin{table} \label{regionstable}
\caption{IRAC Coverage of the Cha~II Molecular Cloud and Off-Cloud Fields}
\begin{center}
\begin{tabular}{cccc}
\hline\hline
Region & \multicolumn{2}{c}{Central Position$^1$}  & Area$^2$  \\
  & R. A. (J2000)  &  Dec (J2000) &  (deg$^2$) \\
  & h min sec & $^\circ$ ~ $\arcmin$ ~ $\arcsec$  &  \\
\hline\hline
ChaII-1  & 13 01 38.88        & $-78$ 01 11.98    &  0.48 \\
ChaII-2  & 13 01 11.52        & $-77$ 34 12.00    &  0.37 \\
ChaII-3  & 12 54 13.68        & $-76$ 58 48.01    &  0.22 \\
OC-1     & 13 37 54.00        & $-75$ 58 48.01    &  0.05 \\
OC-2     & 13 38 02.64        & $-77$ 58 48.01    &  0.05 \\
OC-3     & 13 38 15.36        & $-79$ 58 48.01    &  0.05 \\
OC-4     & 11 58 02.16        & $-77$ 30 36.01    &  0.05 \\
OC-5     & 11 58 07.68        & $-78$ 30 36.01    &  0.05 \\
OC-6     & 11 58 26.16        & $-81$ 00 36.01    &  0.05 \\
\hline\hline
\multicolumn{4}{l}{$^1$of IRAC 3.6 $\&$ 5.8$\mu$m grids}     \\  
\multicolumn{4}{l}{$^2$Area covered by all four IRAC bands}
\end{tabular}
\end{center}
\end{table}

\begin{table}
\caption{Detection of sources with S/N $\ge 7\sigma$ toward Chamaeleon II
  (total numbers). }\label{detect1}
\begin{tabular}{lll}\hline\hline
                                                  & On cloud & Off cloud \\
Detection in at least one IRAC band               &  69848   & 18483 \\
Detection in all 4 IRAC bands                     &   6476   & 1562 \\
Detection in 3 IRAC bands                         &   5460   & 1424 \\
Detection in 2 IRAC bands                         &  27506   & 7017 \\
Detection in 1 IRAC band                          &  30406   & 8480 \\[1.0ex]
Detection in 2MASS only$^{a}$                     &     67   & 0 \\
Detection in IRAC only                            &  65139   & 17222\\
Detection in 4 IRAC bands and not 2MASS$^{a}$     &   2179   & 462 \\\hline
\emph{Excluding extended sources:} & \\
Four band detections                              & 6284     & 1496 \\
Four band detections with 2MASS association$^{a}$ & 4229     & 1086 \\
Detected in IRAC1+2 and 2MASS$^{a}$               & 4621     & 1248 \\ \hline
\end{tabular}

$^{a}$A source is counted as detected by 2MASS if it has a S/N of at
least 10 in both $H$ and $K_s$.
\end{table}

\begin{table}
\caption{Detection of sources toward Chamaeleon II (per band).}\label{detect2}
\begin{tabular}{lllll} \hline\hline
                                                      & 3.6~$\mu$m     & 4.5~$\mu$m     & 5.8~$\mu$m     & 8.0~$\mu$m     \\ \hline
Detections with                                       &                & 
        &                &          \\
$\ldots$S/N of at least 7                             &  61769         &    46221       &    11583       &     8219 \\
$\ldots$S/N of at least 10                            &  55742         &    39034       &     8204       &     5992 \\
$\ldots$S/N of at least 15                            &  39527         &    24636       &     4979       &     4248 \\
Final sample (excluding extended sources)             &  60601         &    45935       &    11513       &     8003 \\
$\ldots$2MASS ass. (S/N at least 10 in $H$ and $K_s$) &   4633         &     4642       &     4443       &     4453 \\\hline
\end{tabular}
\end{table}

\begin{table}
\caption{Statistics of YSO candidates }\label{ysotab}
\begin{tabular}{lllll}\hline\hline
  Quantity       & Cha~II\tbn{a}    & OC\tbn{a}	& SWIRE\tbn{b} 
& New\tablenotemark{c}\\ \hline
Area (deg$^2$)         & 1.04		& 0.3	& 5.3	& 1.04 \\
YSO candidates         &  40          	&  5	& 59	& 24	\\
YSOc's per sq. degree  &  38.5        	& 17	& 11.6	& 23.1 \\
Class I                &  11 (27.5\%) 	& 4	& 3.7	& 2 (8\%) \\
``Flat spectrum''      &   8 (20.0\%) 	& 0	& 5.9	& 0 (0\%) \\
Class II               &  17 (42.5\%) 	&1	& 0.8	& 19 (79\%) \\
Class III              &   4 (10.0\%)   & 0	& 1.2	& 3 (13\%)  \\ \hline
\end{tabular}
\tablenotetext{a}{Based on 2005 criteria for YSO candidates; actual numbers
for classes}
\tablenotetext{b}{Based on 2005 criteria for YSO candidates; numbers
for classes are noralized to area of Cha~II}
\tablenotetext{c}{Based on new criteria for YSO candidates (see discussion 
in \S 6)}
\end{table}

\clearpage

\clearpage
\clearpage
\resizebox{\hsize}{!}{\includegraphics{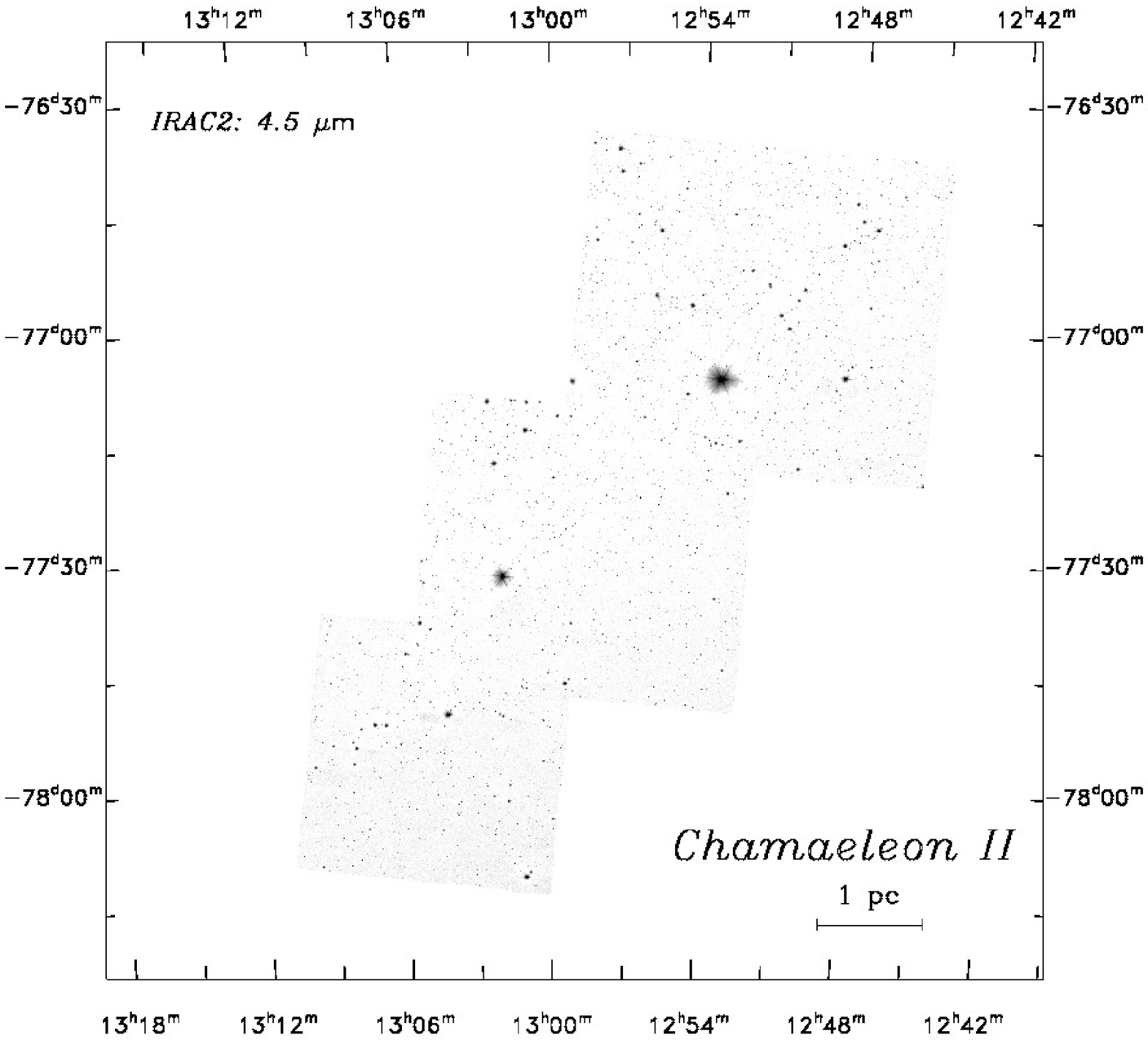}}
\clearpage
\resizebox{\hsize}{!}{\includegraphics{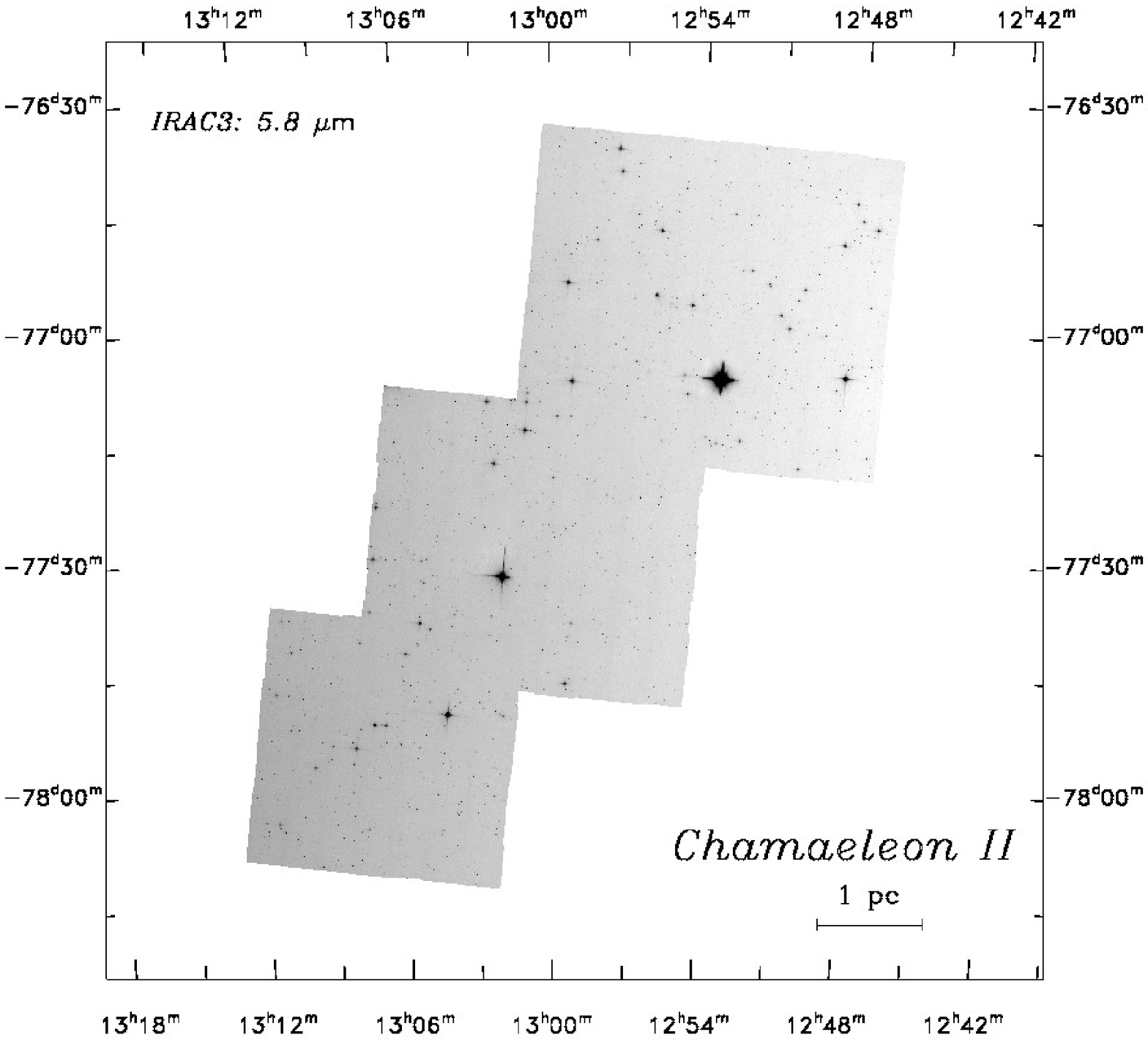}}
\clearpage
\resizebox{\hsize}{!}{\includegraphics{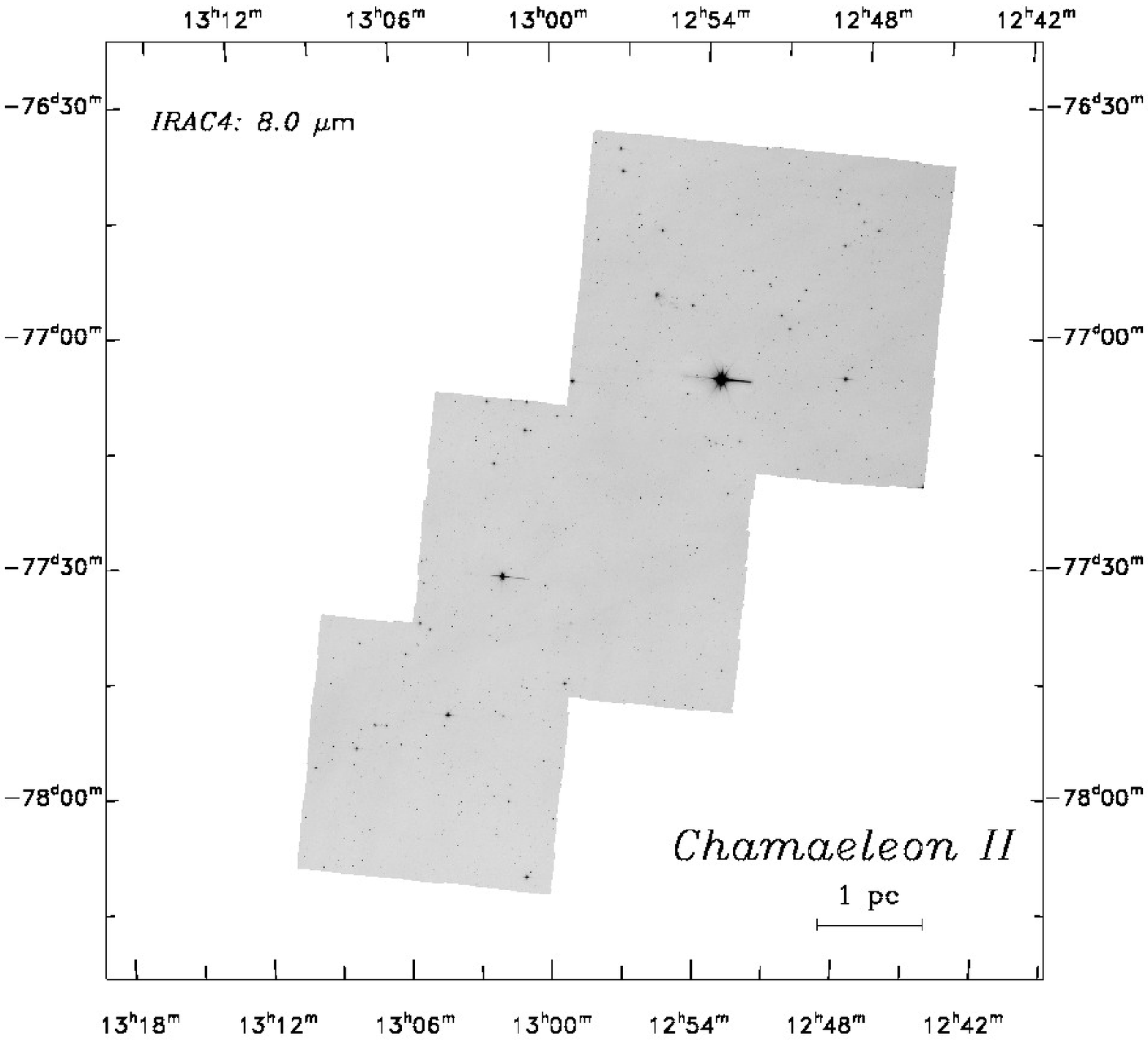}}
\clearpage

\begin{figure}
\includegraphics[scale=0.9]{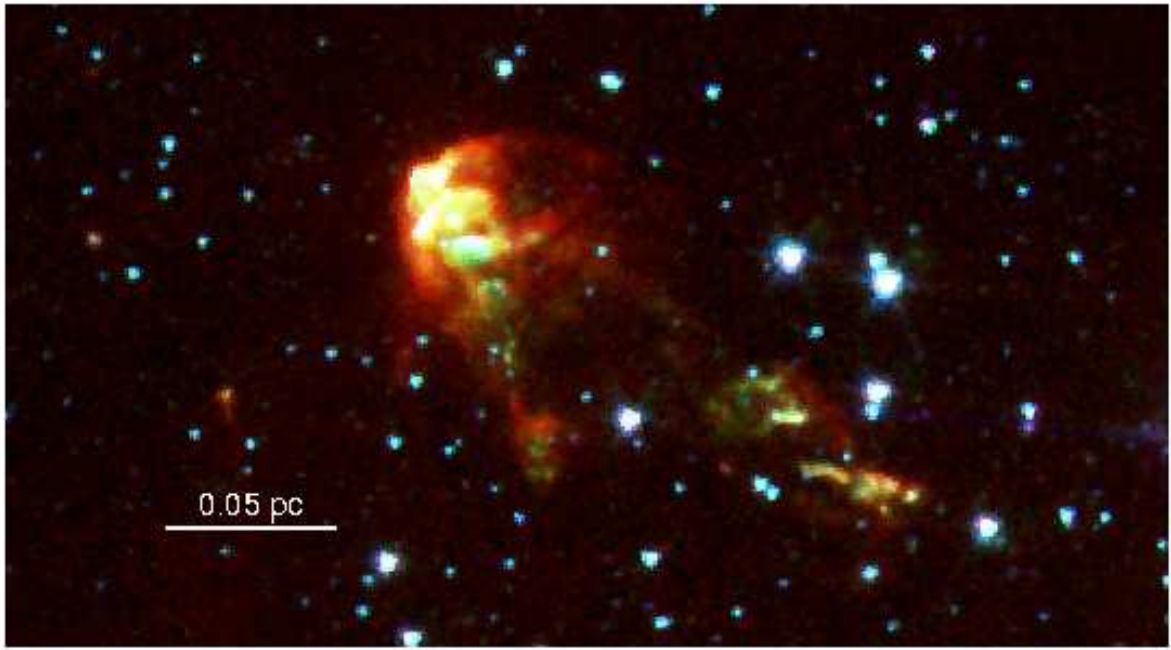}
\figcaption{\label{rgb_hh54} Color composite image of the interesting outflow
HH~54. The coding is 3.6 \micron\ as blue, 4.5 \micron\ as green, and
8.0 \micron\ as red.
}
\end{figure}

\end{document}